\documentclass[fleqn,twoside]{article}
\usepackage[headings]{espcrc2}

\readRCS
$Id: espcrc2.tex,v 1.2 2004/02/24 11:22:11 spepping Exp $
\usepackage{graphicx}
\usepackage[figuresright]{rotating}

\newcommand{\AmS}{{\protect\the\textfont2
  A\kern-.1667em\lower.5ex\hbox{M}\kern-.125emS}}

\title{Algorithm for muon electromagnetic shower reconstruction}

\author{S. Mangano\address[MCSD]{NIKHEF,
        Kruislaan 409, 1098 SJ Amsterdam, The Netherlands}
        on behalf of the ANTARES collaboration}

\runtitle{Roma International Conference on Astro-Particle physics}
\runauthor{Muon electromagnetic shower reconstruction}

\begin{document}

\begin{abstract}
The ANTARES neutrino telescope is presently being built in the
Mediterranean Sea at a depth of 2500 m. The primary aim of the experiment is the
detection of high energy cosmic muon neutrinos, which are
identified by the muons that are produced in charged current
interactions. These muons are detected by measuring the
Cerenkov light which they emit traversing the detector. Sometimes
a high momentum muon produces electromagnetic showers. The subject of 
this paper is a method to
reconstruct these showers which includes several steps: an
algorithm for the fit of the muon track parameters, 
preselection of detected photons belonging
to a shower, and  
a final fit with the preselected detected photons 
to calculate the electromagnetic shower position.
Finally a comparison between data obtained with that part of the detector 
that is currently in operation and simulations is presented.

\vspace{1pc}
\end{abstract}

\maketitle

\section{Introduction}
The ANTARES \cite{Antares} detection principle relies 
on the observation of 
Cerenkov (CK) light emitted by relativistic charged particles in water.  
The emitted CK photons are detected by a  
3-dimensional grid of 900
photomultiplier tubes (PMTs) arranged in 12 detection lines 
at a depth of 2500 m
in the Mediterranean Sea. The detector will be completed in early 2008.
Since the end of January 2007, the detector has 
5 operating detection lines. Amongst others first data of 
atmospheric downward going muons have been measured, 
which are up to $500$ m long tracks with  
energies larger than $100$ GeV.
This gives the unique opportunity to study the 
average number of showers per track length for 
high energy muons passing through sea water.

The muon track is reconstructed from the arrival times of the CK light
detected by the PMTs, whose positions are known. 
The muon energy loss mechanism in water 
are ionization, \mbox{$e^+e^-$ pair} production,
bremsstrahlung, and photonuclear interactions. 
Above several \mbox{hundred GeV} the muon energy loss is dominated 
by the latter three effects \cite{PDG}, which are discrete and are characterized 
by large energy fluctuations. Furthermore, 
the average muon energy loss due to these effects is
proportional to the energy of the muon.

In the following a method to reconstruct discrete
electromagnetic (EM) showers is presented, 
which gives additional information about a muon track
(e.g. the shower multiplicity) as compared to 
existing algorithms only measuring the direction and position of 
the muon track.

\section{Identification of muon EM showers}
The challenge of EM shower reconstruction is to distinguish
CK photons from EM shower photons.
The EM shower light has two key characteristics:
it is produced in one point on the muon path and it arrives, 
in general, delayed compared to the CK light.

\label{identification}
\begin{figure}[htb]
\vspace{3pt}
\includegraphics[width=0.45\textwidth,height=0.32\textwidth,angle=0]{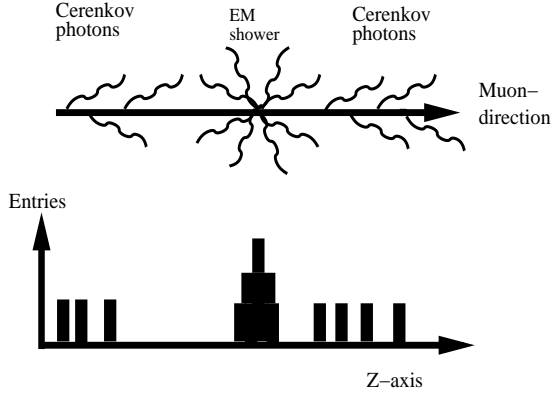}
\vspace{-5pt}
\caption{\textit{A muon emits Cerenkov photons continuously on its trajectory in water with sometimes a discrete electromagnetic shower (upper panel). The emission points of the photons from the muon trajectory are calculated from the measured \mbox{3-dimensional} position and arrival time in the PMTs. These emission points are presented by a one-dimensional histogram with its Z-axis aligned with the muon axis. In the histogram the peaks identify the electromagnetic shower position (lower panel).}}
\label{fig:shower}
\end{figure}
The first characteristic distinguishing CK light from EM
shower light is illustrated in \mbox{Fig. \ref{fig:shower}.} 
The CK photons are emitted continuously along the muon path, whereas 
EM showers are emitted stochastically and discretely.
Furthermore, the number of photons from EM showers is proportional 
to the deposited energy. 
By projecting the 3-dimensional position and arrival time
of all detected photons onto the muon axis, the identification of EM 
showers is reduced to a one-dimensional problem. 
This transformed information 
is presented by a one-dimensional histogram, in which the 
EM shower position appears as a peak. So, the peak 
indicates where on the muon trajectory the EM shower is produced.

\section{Algorithm for shower reconstruction}
The algorithm consist of three steps using 
digitized PMT signals which are referred to as hits\footnote{ 
A hit occurs when a PMT signal crosses a threshold 
corresponding to 0.3 of a single photon-electron amplitude.}. 
The first step selects muon candidates. Then,
hits are preselected
which originate from one distinct shower. 
The third step is a $\chi^2$-minimization yielding 
the space-time position of the shower within the 
preselected hits.

\subsection{Selection of shower hits}
\label{step2}
This section presents the mathematical framework which is schematically 
illustrated by Fig. \ref{fig:shower}.   

\begin{figure}[htb]
\vspace{3pt}
\includegraphics[width=0.45\textwidth,height=0.22\textwidth,angle=0]{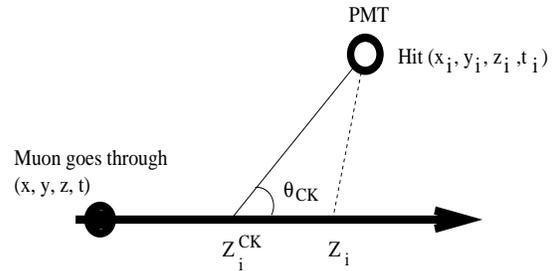}
\vspace{-5pt}
\caption[]{\textit{{Schematic view of the detection of the Cerenkov light and the randomly emitted shower light. The thick line represents the muon, the thin line the path of the Cerenkov light and the thin dotted line the path of the shower light. The muon goes through point $(x, y, z, t)$. The Cerenkov light is emitted at an angle $\theta_{CK}$ with respect to the muon track at point $Z_i^{CK}$ and is detected by a PMT as a hit in point $(x_i, y_i, z_i, t_i)$. The randomly emitted shower light is emitted at point $Z_i$ and is detected by the same PMT, but later in time.}}}
\label{fig:calc}
\end{figure}

Muon events are selected by using an existing muon reconstruction algorithm
described in \cite{Ronald}, which provides an estimate of 
the direction and position of the muon at some fixed time.
For all hits in one muon event, the hit time is considered 
relative to the expected arrival time 
of a direct CK photon from the reconstructed muon track. 
This calculation can be simplified by a rotation 
of the coordinate system. The $M$ hits 
$\vec{x_i'}=(x_i',y_i',z_i',t_i')$ with $i=1,....,M$ of an event
are rotated by the rotation matrix: 
\begin{displaymath}
R(\theta,\phi)= 
\left( \begin{array}{cccc}
\cos \theta \cos \phi & \cos \theta \sin \phi & -\sin \theta & 0\\
 -\sin \phi           & \cos \phi             &      0       & 0\\
\sin \theta \cos \phi & \sin \theta \sin \phi &  \cos \theta & 0\\
0                     &  0                    &      0       & 1\\        
\end{array} \right),
\end{displaymath}
where $\theta$ and $\phi$ are given by the muon direction. The situation 
after the rotation \mbox{$\vec{x_i}=R(\theta,\phi) \vec{x_i'}$} 
of the $i$-th measured hit $\vec{x_i}=(x_i,y_i,z_i,t_i)$  
as well as the rotated muon space-time coordinates $(x,y,z,t)$ 
is presented in Fig. \ref{fig:calc}. 

The expected CK arrival time ($t^{CK}_i$) of the  
photon is calculated by (see \mbox{Fig. \ref{fig:calc})}:
\begin{equation}
  t^{CK}_i = t + \frac{1}{c} \Big(z_i-z - \frac {r_i}{\tan \theta_{CK}}\Big) + \frac{n}{c} \frac{r_i}{\sin \theta_{CK}},
\end{equation}
where $c$ is the speed of light in vacuum, $n$ is the refraction index of water, $\theta_{CK}$ is the CK angle
and $r_i$ is the perpendicular distance between the muon trajectory and the PMT given by 
$r_i=\sqrt{ (x_i-x)^2 + (y_i-y)^2}$.

\subsubsection{Calculation of the photon emission point from the muon}
\label{zcalc}

In order to suppress random background hits, a selection is made 
of hits for which the difference 
between the calculated arrival time $t^{CK}_i$ of CK photons and the measured arrival time 
is in the interval \mbox{$-t_{min} < t_i-t^{CK}_i < t_{max}$.}
Furthermore, this interval is subdivided in two unequal intervals, 
because photons from EM showers arrive in general 
later than CK photons. The early time interval contains 
CK photons and is given by $ |t_i-t^{CK}_i| < t_{min}$. 
The late time interval $t_{min} < t_i-t^{CK}_i < t_{max}$ contains mostly EM shower photons.
The value for $t_{min}$ is given by the timing resolution of the PMT 
and the dispersion of light in water, 
whereas $t_{max}$ is defined by a value where the number of signal hits 
approaches the background level. 

The emission location of photons  $Z_i^{CK}$ from the early time interval is calculated under the CK angle assumption and is given by (see Fig. \ref{fig:calc}):
\begin{equation}
Z_i^{CK}=z_i- z - \frac {r_i}{\tan \theta_{CK}}.
\label{equ:cvpos}
\end{equation}
For the late time interval the unknown photon emission location $Z_i$ is given by (see \mbox{Fig. \ref{fig:calc})}:
\begin{eqnarray}
\lefteqn{ t_i = t + \frac{Z_i-z}{c} + {} } \nonumber \\
& & {} + \frac{n}{c} \sqrt{ (x_i-x)^2 + (y_i-y)^2 + (z_i- Z_i )^2 },
\label{equ:showertime}
\end{eqnarray}
where the second term is the time for a muon moving 
at speed of light to reach the 
point  $Z_i$ where the light is emitted,  while 
the third term
is the time needed for the light in water to travel from $Z_i$ to the PMT.  
Eq. \ref{equ:showertime} corresponds to a quadratic equation with two distinct 
solutions\footnote{The late time interval ensures that two real solutions exist.} represented by $Z_i^{+}$ and $Z_i^{-}$.

\subsubsection{Peak finder}
\label{zcalc}
\begin{figure}[tb]
\vspace{-3pt}
\includegraphics[width=0.5\textwidth,height=0.515\textwidth,angle=0]{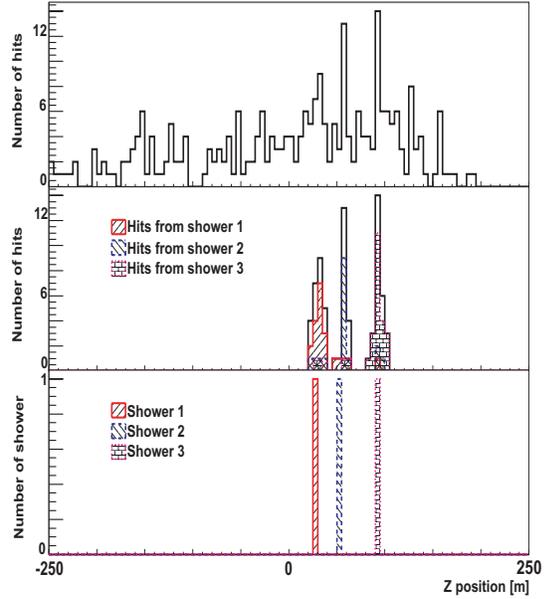}
\vspace{-35pt}
\caption[Sc]{\textit{The upper panel represents all reconstructed emission points of the photons emitted by an atmospheric muon. They are calculated from the measured hit information after a full simulation of the detector response to an atmospheric muon event with backgrounds noise. The middle panel shows the hits which are selected from the algorithm after the peak finder step (see text). Only statistically relevant peaks have been selected. Furthermore, hits produced from each generated shower are represented by a different hatching, emphasizing that the purity of each peak is high. The lower panel displays the three generated shower positions on the muon axes.}}
\label{fig:peaks}
\vspace{-5pt}
\end{figure}

All calculated $Z_i^{CK}$, $Z_i^{+}$ and $Z_i^{-}$ positions 
are presented by a one-dimensional histogram. The upper panel 
of Fig. \ref{fig:peaks} shows an example of a
full simulation of the detector
response to an atmospheric muon event with background noise.

These  $Z_i^{CK}$, $Z_i^{+}$ and $Z_i^{-}$ positions are used 
by a one-dimensional peak finder algorithm.
It selects peaks with amplitudes higher than a threshold identifying 
only statistically relevant peaks.
The peak finder takes into account that if the two solution $Z_i^{+}$ and $Z_i^{-}$ are 
used in different peaks, the solution in 
the smaller peak is eliminated. 
The result of the peak finder is shown in the middle panel
of Fig. \ref{fig:peaks}.  
In addition hits produced from a particular shower 
are represented by a different hatching. The non-hatched hits are 
either produced by CK light or background noise. 

In the lower panel of Fig. \ref{fig:peaks} the generated position
of the three showers are also presented. The agreement between the middle and lower panel shows that the  
selected peaks in this event have a high hit-purity and the algorithm succeeds in properly reconstructing the shower 
position on the muon axes.

\subsection{Reconstruction of the shower position}
\label{step3}
The last step of the algorithm is to calculate the 
EM shower space-time position
under the assumption of a point-like shower source for each peak.
For this, a first estimate of the shower space-time position 
$\vec{x_s}=(x_s,y_s,z_s,t_s)$  as explained in \cite{Hartmann} is used.
Then for a peak with $N$ hits $\vec{x_j'}=(x_j',y_j',z_j',t_j')$ 
a $\chi^2$ with a fixed $\sigma$ can be introduced: 
\begin{eqnarray}
\lefteqn{ \chi^2=\sum_{j=1}^{N} \frac{1} {\sigma^2} \cdot \Big(t_j'-t_s- {} } \nonumber \\
 & & {}\frac{n}{c}\sqrt{(x_j'-x_s)^2+(y_j'-y_s)^2+(z_j'-z_s)^2} \Big)^2
\label{equ:chi2}
\end{eqnarray}

If necessary the $\chi^2$ is minimized by 
removing hits until among all possible permutations 
of the hits used to calculate 
the shower position a satisfactory $\chi^2$ 
normalized to the number of freedom is found.

The result of the algorithm is that for each muon candidate the 
shower multiplicity can be calculated. For each shower candidate 
different quantities can be evaluated: e.g.
the position of the shower, the closest approach 
between shower position and muon track and the number of hits per shower.

\section {Conclusion}
\label{conclusion} 
\begin{figure}[tb]
\vspace{3pt}
\includegraphics[width=0.5\textwidth,height=0.5\textwidth,angle=0]{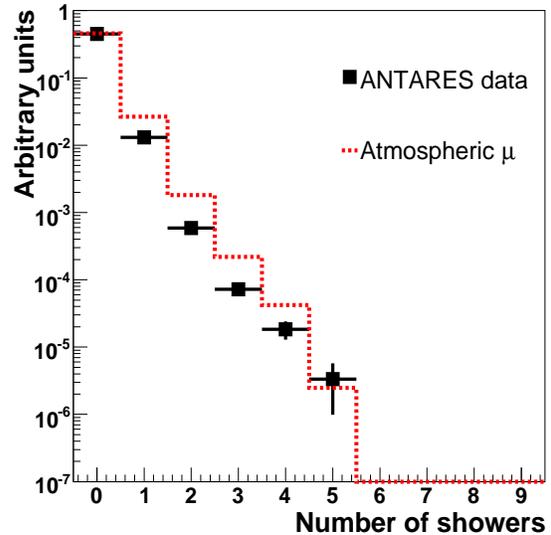}
\vspace{-30pt}
\caption[Sc]{\textit{Distribution of the number of reconstructed atmospheric muon shower candidates for both data (points) and simulation (dotted line).}}
\label{fig:datamcshowermore}
\vspace{10pt}
\end{figure}

The performance of the reconstruction algorithm for EM showers induced by muons 
has been validated in a sample of simulated multiple atmospheric muon events with 
background noise. Next, the algorithm was applied to data obtained with ANTARES.    
The results of the shower multiplicity for down going muon events is
shown in Fig. \ref{fig:datamcshowermore}.
The agreement between data and simulation is satisfactory.
For the used set of selection cuts, the one shower muon 
events constitute about 5\% of all muon events. 

In summary a method to reconstruct EM showers from a muon has 
been developed and applied for the first time in ANTARES.
The essential element of the algorithm is that the selection of shower hits is 
reduced to a one-dimensional problem.
Improvements of the algorithm as well as a more extensive application to data are subject 
of the future work.

\end{document}